# Electronic Ground State of Higher Acenes

*De-en Jiang[1,*] and Sheng Dai[1,2]*

[1]Chemical Sciences Division and [2]Center for Nanophase Materials Sciences, Oak Ridge National Laboratory, Oak Ridge, Tennessee 37831

jiangd@ornl.gov

Electronic ground state of higher acenes

*To whom correspondence should be addressed. E-mail: jiangd@ornl.gov. Phone: (865)574-5199. Fax: (865) 576-5235.

We examine the electronic ground state of acenes with different number of fused benzene rings (up to 40) by using first principles density functional theory. Their properties are compared with those of infinite polyacene. We find that the ground state of acenes that consist of more than seven fused benzene rings is an antiferromagnetic (in other words, open-shell singlet) state, and we show that this singlet is not necessarily a diradical, because the spatially separated magnetizations for the spin-up and spin-down electrons increase with the size of the acene. For example, our results indicate that there are about four spin-up electrons localized at one zigzag edge of 20-acene. The reason that both acenes and polyacene have the antiferromagnetic ground state is due to the zigzag-shaped boundaries, which cause π-electrons to localize and form spin orders at the edges. Both wider graphene ribbons and large rectangular-shaped polycyclic aromatic hydrocarbons have been shown to share this antiferromagnetic ground state. Therefore, we demonstrate that the π-electronic structures of higher acenes and ployacene are still dictated by the zigzag edges, and our results provide a consistent description of their electronic ground state.

Keywords: acene, polyacene, electronic structure, ground state, density functional theory, zigzag edge

## 1. Introduction

Acenes have attracted great interest from a wide spectrum of researchers.[1-6] These linearly fused benzene rings (see Figure 1a) provide interesting electronic properties due to the conjugated π-electron system. Especially, pentacene has been widely used in molecular electronics.[7-19]

Acenes belong to polycyclic aromatic hydrocarbons (PAH). Great progress has been made towards the synthesis of large PAHs[20] since the publication of Clar's two volumes on polycyclic hydrocarbons.[21] In the books, Clar pointed out heptacene's enormous reactivity that rendered it impossible to obtain heptacene in a pure form. Clar also described early unsuccessful efforts to synthesize octacene, nonacene, and undecacene. Only very recently was heptacene-containing single crystal obtained,[22] and higher acenes remain elusive. So does the corresponding polymer, polyacene.[23]

Several theoretical studies have attempted to understand the electronic structures of higher acenes and explain their high reactivity. Houk and co-workers predicted a triplet ground state for acenes with n (the number of fused benzene rings) > 8,[24] using unrestricted B3LYP (UB3LYP) for the triplet but restricted B3LYP for the singlet (both with 6-31G* basis). Later, Bendikov et al. showed that the ground state is an open-shell singlet for n > 6, by using UB3LYP (with 6-31G* basis) also for the singlet, and claimed that this open-shell singlet represents a diradical.[25,26] Because a diradical is a molecule with two unpaired electrons,[27] their claim implies that there are one up and one down spins in acenes with n > 6. In a recent study, dos Santos[28] reported a higher spin ground state, a quintet, for n=20-23, also based on UB3LYP/6-31G*. More recently, Chan and coworkers used a density matrix renormalization group algorithm and studied acenes with n=2-12. They found that the ground states for longer acenes are polyradical singlets.[29]

In the meantime, there has been much interest in zigzag-edged graphene nanoribbons (ZGNR),[30-39] which can be viewed as parallel polyacetylene chains cross-linked together (the number of parallel polyacetylene chains is used as an index to characterize the width of a ZGNR). Acenes and polyacene also share this structural feature of zigzag edges. We and others[30,40] have predicted that infinitely long ZGNRs with a width index of four or higher have an antiferromagnetic (AFM) ground state and each edge carbon atom has a finite local magnetic moment. This result demands that the number of unpaired electrons in a finite-sized ribbon increase with the ribbon length as the properties of acenes approach those of polyacene. We have observed this trend in finite-sized ZGNRs with a width index higher than two.[41] Because polyacene can be viewed as a ZGNR with a width index of two, one would expect that acenes will also have an increasing number of unpaired electrons with the number of fused rings if polyacene has an AFM ground state as well, thereby indicating that the diradical concept would be inaccurate for higher acenes. To show that this is indeed the case, in this paper we employ a spin-polarized density functional theory (DFT) method, which is less prone to spin contamination,[42] to first address the electronic ground state of polyacene and then that of acenes.

## 2. Method

DFT calculations using the Vienna Ab Initio Simulation Package were performed,[43,44] based on planewave bases and periodic boundary conditions and within the generalized-gradient approximation for electron exchange and correlation.[45] Projector-augmented wave method was used within the frozen core approximation to describe the electron-core interaction.[46,47] Supercell models were employed; i.e., an acene molecule was put in a large box. The flat molecule was placed in the $xy$-plane (see Figure 1a). The $y$ and $z$ dimensions of the boxes were fixed at 15 and 10 Å, respectively, while the $x$ dimension increased from 23 Å for pentacene to 116 Å for 40-acene. Only the Γ-point was used to sample the Brillouin zone for the molecules. For polyacene, the unit cell is shown in Figure 1b. The $x$-dimension (i.e., the repeating length of the polymer) was optimized to be 2.461 Å, and 49 irreducible k-points were used to sample the Brillouin zone.

A kinetic energy cutoff (450 eV) was used, and all atoms in the supercell were allowed to relax and the force tolerance was set at 0.025 eV/Å. Full relaxation of magnetization was performed for spin-polarized calculations.

## 3. Results and Discussion

We started our investigation with polyacene. Although synthesis of polyacene has not been reported, numeral theoretical examinations exist and conflicting conclusions have been reached.[27,48-54] Detailed discussion of previous literature about polyacene is documented in Ref. 57. Because we will address the electronic structures of polyacene in comparison with other polymers in greater detail in a forthcoming publication, here we only present our major results for polyacene and focus discussion on acenes.

We find that the antiferromagnetic (AFM) state is the ground state of polyacene with an optimized structure of a symmetric configuration ($R_1=R_2=1.406$ Å, $R_3=1.459$ Å). The energy is about ~10 meV/cell lower than the nonmagnetic (NM) phase. The fact that the AFM phase is the ground state for polyacene agrees with what we and others have found for the ZGNRs and rectangular PAHs.[38,40,41] This AFM ground state has been attributed to the zigzag edges and the electron-electron interaction.[30] Fujita et al. have shown that an infinitesimal on-site Coulomb repulsion in the Hubbard model with unrestricted Hartree-Fock approximation can cause an AFM order in a ZGNR with a width index of 10.[30] Because polyacene can be viewed as the narrowest ZGNR (if one excludes polyacetylene), it is understandable that polyacene shares a common ground state with other wider ZGNRs. Although an ferromagnetic (FM) state has been found for wider ZGNRs,[40,55] we did not find an FM phase for polyacene.

With the results of infinite polyacene in mind, we can now discuss acenes. Figure 2 shows how the relative energetics change with the size of the oligomer from 5 to 20. One can see that the AFM state[56] emerges as a ground state when n is greater than 7. This AFM ground state is consistent with what we have found for polyacene, because acenes should resemble polyacene when n is large enough. At n=7, the AFM and nonmagnetic (NM)[57] phases are very close in energy. When n is smaller than 7, the NM phase is the ground state. The AFM state has high chemical reactivity because unpaired electrons are piled up at the edges and yield partial radical characters at the edge carbon atoms.[41] One can see this from the spin density magnetization in Figure 3 for n=10. Therefore, the appearance of the AFM phase can be used an indicator of acenes' chemical reactivity. The fact that heptacene-containing single crystals have been obtained only recently and that acenes with more than 7 fused rings have not been synthesized supports the use of the AFM state as a reactivity indicator.

In addition to the AFM state, we investigated the ferromagnetic (FM)[58] triplet state and found that its energy is always higher than the corresponding AFM state and oscillates above and below the NM state. We also investigated quintet for all the acenes and found that their energies are all higher than the triplet.

The relative energetic trend that we have obtained for acenes agrees quite well with a recent hybrid functional study with localized basis sets (UB3LYP/6-31G*) by Bendikov et al.[25,26] They found that the open-shell singlet (corresponding to our AFM state) configuration is the ground state for n=6-10. Our results for n=20 are in contrast with dos Santos's finding[28] that acenes (n=20-23) have a quintet ground state (based on UB3LYP/6-31G*). Because the open-shell singlet was not specifically examined in his work, whether its energy will be lower than his quintet state is not clear from his work.

In their paper, Bendikov et al. further claimed that the open-shell singlet is a diradical for acenes with n > 6, thereby implying that acenes have one spin-up and one spin-down unpaired electrons in their ground state. To demonstrate that this diradical concept is inaccurate for higher acenes, we show the number of unpaired spin-up electrons for several n-acenes by integrating their spin density magnetization in Table 1. One can see that the total number of unpaired spin-up electrons increase with the size of the acene and is greater than one even for n=10. In addition, one can see that the average spin-up moments per benzene ring slowly increases with n from 10 to 20 to 40 and approaches that of polyacene. Therefore, our results provide a consistent picture of how finite-sized n-acene approaches infinite polyacene with n.

Now we analyze how the unpaired spins are distributed among frontier orbitals, using 20-acene as an example. Figure 4 displays the Kohn-Sham orbital spectrum near the Fermi level for 20-cene. We singled out the three highest occupied spin orbitals (HOSO) for the spin-up electron. By decomposing the total charge density and spin magnetization, we obtained the contributions of HOSO, HOSO2, and HOSO3 to the total magnetization. They are 0.909, 0.739, and 0.456 e, respectively. Together, these three orbitals contribute more than two unpaired spin-up electrons, while a diradical picture would limit the total number to one. This result clearly indicates that the diradical description is inaccurate for higher acenes. We visualize the electron densities for the three HOSOs in Figure 5. One can see the localizations of spin densities at the edges. This localization is due to the zigzag edges and has been found in ZGNRs and rectangular PAHs.[30,31,41] The data in Table 1 and our analysis of 20-acene indicate that the ground state (i.e., the AFM state) of higher acenes (n > 7) is generally not a diradical, because they tend to have more than one unpaired spin-up electrons (and more than two unpaired electrons) and the number of unpaired electrons also increases with the size.

In a recent study employing a density matrix renormalization group (DMRG) algorithm for acenes with n=2-12, Chan and coworkers reached the same conclusion that longer acenes show singlet polyradical character with antiferromagnetically coupled electrons. Because their method takes into account electron correlation fully in the π-valence space, their level of theory is considerably higher than the present one. Thus, the fact that we reached the same conclusion shows that the computationally less intensive, standard spin-polarized DFT method employed in the present work describes quite accurately the ground state of π-electrons in acenes.

The C-C bond length alternations in $sp^2$ networks have been used to understand the aromaticity of the networks.[59] Figure 6 displays the C-C bond lengths along the upper zigzag edge (which is mirror-symmetric with the lower edge) for the AFM and NM states. One can clearly see large variations at the two end benzene rings, indicating formation of two conjugated double bonds in the two end rings. Inner rings have smaller C-C bond variations, especially for the AFM state. In fact, the C-C bond lengths in the middle eight rings for the AFM state vary only slightly and level at ~ 1.406 Å, which equals to the corresponding C-C bond length in polyacene. This comparison of bond lengths demonstrates that the middle rings resemble polyacene. Analysis of local magnetization also support this conclusion in that local magnetic moments on zigzag edge carbon atoms in the middle rings are similar to that of polyacene. We have shown high chemical reactivity at the edges of infinite ZGNRs previously.[40,60] Therefore, inner rings of higher acenes, resembling infinite polyacene, should be more reactive than the outer ones, in agreement with a previous analysis.[61]

### 4. Summary and conclusions

The electronic ground states of acenes with different number of fused benzene rings (n up to 40) and polyacene have been studied with first principles spin-polarized density functional theory. The ground states of higher acenes (n > 7) and polyacene are found to be antiferromagnetic. This antiferromagnetic ground state is due to the zigzag-shaped boundaries, which cause π-electrons to localize and form spin orders at the edges. Unlike previously proposed, we show that the antiferromagnetic state is not necessarily a diradical for acenes. The reason is that the edge-localized magnetizations for the spin-up and spin-down electrons increase with the size of the acene. Therefore, longer acenes will have more than two unpaired electrons.

Acknowledgement

This work was supported by Office of Basic Energy Sciences, U.S. Department of Energy under Contract No. DE-AC05-00OR22725 with UT-Battelle, LLC. This research used resources of the National Energy Research Scientific Computing Center, which is supported by the Office of Science of the U.S. Department of Energy under Contract No. DE-AC02-05CH11231.

Figure 1. (a) Acene with n fused benzene rings, designated as n-acene; (b) polyacene (unit cell is enclosed in brackets).

Figure 2. Energetics of the AFM and FM (triplet) states relative to the NM state of n-acene vs. the number of fused benzene rings (n).

Figure 3. Isosurfaces of spin density magnetization ($\rho_\uparrow - \rho_\downarrow$) for the AFM phase of 10-acene. Dark and light isosurfaces are 0.075 and -0.075 e/Å$^3$, respectively.

Figure 4. Kohn-Sham orbital energies near the Fermi level for spin up and spin down electrons of 20-acene. HOSO, HOSO2 and HOSO3 are the three highest occupied spin orbitals.

Figure 5. Isosurfaces of orbital-decomposed charge density for spin up electrons of 20-acene. Only the three highest occupied spin orbitals are shown. Isosurface values are at 0.02 e/Å$^3$.

Figure 6. C-C bond lengths along the upper zigzag edge of 20-acene.

Table 1. Sum of spin-up moments of spin density magnetization ($\rho_\uparrow - \rho_\downarrow$) ($M_{\text{up-AFM}}$) and averaged magnetic moment per benzene ring ($M_{\text{edge-AFM}}$) for the antiferromagnetic phase of polyacene and n-acenes.

| System | $M_{\text{up-AFM}}$ ($\mu_B$)[a] | $M_{\text{edge-AFM}}$ ($\mu_B$) |
|---|---|---|
| Polyacene | 0.226 | 0.226 |
| 7-acene | 0.96 | 0.137 |
| 10-acene | 2.030 | 0.203 |
| 15-acene | 2.839 | 0.189 |
| 20-acene | 4.103 | 0.205 |
| 40-acene | 8.315 | 0.208 |

[a] Per unit cell for polyacene and per molecule for acenes.

References and Notes

(1) Mondal, R.; Shah, B. K.; Neckers, D. C. *J. Am. Chem. Soc.* **2006**, *128*, 9612.

(2) Kadantsev, E. S.; Stott, M. J.; Rubio, A. *J. Chem. Phys.* **2006**, *124*.

(3) Knupfer, M.; Berger, H. *Chem. Phys.* **2006**, *325*, 92.

(4) Xie, H. M.; Wang, R. S.; Ying, J. R.; Zhang, L. Y.; Jalbout, A. F.; Yu, H. Y.; Yang, G. L.; Pan, X. M.; Su, Z. M. *Adv. Mater.* **2006**, *18*, 2609.

(5) Lindstrom, C. D.; Muntwiler, M.; Zhu, X. Y. *J. Phys. Chem. B* **2007**, *111*, 6913.

(6) Yamakita, Y.; Kimura, J.; Ohno, K. *J. Chem. Phys.* **2007**, *126*.

(7) Horowitz, G.; Peng, X. Z.; Fichou, D.; Garnier, F. *Synth. Met.* **1992**, *51*, 419.

(8) Dimitrakopoulos, C. D.; Brown, A. R.; Pomp, A. *J. Appl. Phys.* **1996**, *80*, 2501.

(9) Brown, A. R.; Jarrett, C. P.; deLeeuw, D. M.; Matters, M. *Synth. Met.* **1997**, *88*, 37.

(10) Nelson, S. F.; Lin, Y. Y.; Gundlach, D. J.; Jackson, T. N. *Appl. Phys. Lett.* **1998**, *72*, 1854.

(11) Dimitrakopoulos, C. D.; Purushothaman, S.; Kymissis, J.; Callegari, A.; Shaw, J. M. *Science* **1999**, *283*, 822.

(12) Gelinck, G. H.; Geuns, T. C. T.; de Leeuw, D. M. *Appl. Phys. Lett.* **2000**, *77*, 1487.

(13) Afzali, A.; Dimitrakopoulos, C. D.; Breen, T. L. *J. Am. Chem. Soc.* **2002**, *124*, 8812.

(14) Meng, H.; Bendikov, M.; Mitchell, G.; Helgeson, R.; Wudl, F.; Bao, Z.; Siegrist, T.; Kloc, C.; Chen, C. H. *Adv. Mater.* **2003**, *15*, 1090.

(15) Katz, H. E. *Chem. Mat.* **2004**, *16*, 4748.

(16) Tamura, R.; Lim, E.; Manaka, T.; Iwamoto, M. *J. Appl. Phys.* **2006**, *100*, 114515.

(17) Fujimori, F.; Shigeto, K.; Hamano, T.; Minari, T.; Miyadera, T.; Tsukagoshi, K.; Aoyagi, Y. *Appl. Phys. Lett.* **2007**, *90*, 193507.

(18) Iechi, H.; Watanabe, Y.; Kudo, K. *Jpn. J. Appl. Phys. Pt. 1* **2007**, *46*, 2645.

(19) Stoliar, P.; Kshirsagar, R.; Massi, M.; Annibale, P.; Albonetti, C.; de Leeuw, D. M.; Biscarini, F. *J. Am. Chem. Soc.* **2007**, *129*, 6477.

(20) Wu, J. S.; Pisula, W.; Mullen, K. *Chem. Rev.* **2007**, *107*, 718.

(21) Clar, E. *Polycyclic Hydrocarbons*; Academic Press: London, 1964; Vols. 1, 2.

(22) Payne, M. M.; Parkin, S. R.; Anthony, J. E. *J. Am. Chem. Soc.* **2005**, *127*, 8028.

(23) We note that polyacene, acene, and oligoacene are often used exchangeably in the literature. In this paper, we use polyacene to mean the infinite polymer, consisting of linearly fused benzene rings.

(24) Houk, K. N.; Lee, P. S.; Nendel, M. *J. Org. Chem.* **2001**, *66*, 5517.

(25) Bendikov, M.; Duong, H. M.; Starkey, K.; Houk, K. N.; Carter, E. A.; Wudl, F. *J. Am. Chem. Soc.* **2004**, *126*, 7416.

(26) Bendikov, M.; Duong, H. M.; Starkey, K.; Houk, K. N.; Carter, E. A.; Wudl, F. *J. Am. Chem. Soc.* **2004**, *126*, 10493.

(27) Salem, L.; Rowland, C. *Angew. Chem.-Int. Edit.* **1972**, *11*, 92.

(28) dos Santos, M. C. *Phys. Rev. B* **2006**, *74*, 045426.

(29) Hachmann, J.; Dorando, J. J.; Avilés, M.; Chan, G. K.-L. *J. Chem. Phys.* **2007**, *127*, 134309; this paper was published about two months after we had submitted the original manuscript and was added during revision.

(30) Fujita, M.; Wakabayashi, K.; Nakada, K.; Kusakabe, K. *J. Phys. Soc. Jpn.* **1996**, *65*, 1920.

(31) Nakada, K.; Fujita, M.; Dresselhaus, G.; Dresselhaus, M. S. *Phys. Rev. B* **1996**, *54*, 17954.

(32) Enoki, T.; Kobayashi, Y. *J. Mater. Chem.* **2005**, *15*, 3999.

(33) Kobayashi, Y.; Fukui, K.; Enoki, T.; Kusakabe, K.; Kaburagi, Y. *Phys. Rev. B* **2005**, *71*, 193406.

(34) Ezawa, M. *Phys. Rev. B* **2006**, *73*, 045432.

(35) Kobayashi, Y.; Fukui, K.; Enoki, T.; Kusakabe, K. *Phys. Rev. B* **2006**, *73*, 125415.

(36) Kobayashi, Y.; Kusakabe, K.; Fukui, K.; Enoki, T. *Physica E* **2006**, *34*, 678.

(37) Niimi, Y.; Matsui, T.; Kambara, H.; Tagami, K.; Tsukada, M.; Fukuyama, H. *Phys. Rev. B* **2006**, *73*, 85421.

(38) Son, Y.-W.; Cohen, M. L.; Louie, S. G. *Nature* **2006**, *444*, 347.

(39) Son, Y.-W.; Cohen, M. L.; Louie, S. G. *Phys. Rev. Lett.* **2006**, *97*, 216803.

(40) Jiang, D. E.; Sumpter, B. G.; Dai, S. *J. Chem. Phys.* **2007**, *126*, 134701.

(41) Jiang, D. E.; Sumpter, B. G.; Dai, S. *J. Chem. Phys.* **2007**, *127*, 124703.

(42) Jensen, F. *Introduction to Computational Chemistry*; John Wiley and Sons: Chichester, UK, 1999.

(43) Kresse, G.; Furthmüller, J. *Phys. Rev. B* **1996**, *54*, 11169.

(44) Kresse, G.; Furthmüller, J. *Comput. Mater. Sci.* **1996**, *6*, 15.

(45) Perdew, J. P.; Burke, K.; Ernzerhof, M. *Phys. Rev. Lett.* **1996**, *77*, 3865.

(46) Blöchl, P. E. *Phys. Rev. B* **1994**, *50*, 17953.

(47) Kresse, G.; Joubert, D. *Phys. Rev. B* **1999**, *59*, 1758.

(48) Whangbo, M. H.; Hoffmann, R.; Woodward, R. B. *Proc. R. Soc. London A* **1979**, *366*, 23.

(49) Kivelson, S.; Chapman, O. L. *Phys. Rev. B* **1983**, *28*, 7236.

(50) Tanaka, K.; Ohzeki, K.; Nankai, S.; Yamabe, T.; Shirakawa, H. *J. Phys. Chem. Solids* **1983**, *44*, 1069.

(51) O'Connor, M. P.; Wattstobin, R. J. *J. Phys. C* **1988**, *21*, 825.

(52) Kertesz, M.; Lee, Y. S.; Stewart, J. J. P. *Int. J. Quantum Chem.* **1989**, *35*, 305.

(53) Srinivasan, B.; Ramasesha, S. *Phys. Rev. B* **1998**, *57*, 8927.

(54) Raghu, C.; Pati, Y. A.; Ramasesha, S. *Phys. Rev. B* **2002**, *65*, 155204.

(55) Lee, H.; Son, Y. W.; Park, N.; Han, S. W.; Yu, J. J. *Phys. Rev. B* **2005**, *72*, 174431.

(56) The AFM state in our calculations of molecules using a supercell spin-polarized DFT approach corresponds to an unrestricted open-shell singlet state in a molecular orbital calculation with a cluster model.

(57) The NM state in our calculations of molecules using a supercell DFT approach corresponds to a restricted closed-shell state in a molecular orbital calculation with a cluster model.

(58) Here we use the term "ferromagnetic" in a rather loose sense, just to indicate that the spins at the two edges are all up. Strictly speaking, this phase should be called ferrimagnetic because some of the carbon atoms between the two edges are spin down.

(59) Kertesz, M.; Choi, C. H.; Yang, S. J. *Chem. Rev.* **2005**, *105*, 3448.

(60) Jiang, D. E.; Sumpter, B. G.; Dai, S. *J. Phys. Chem. B* **2006**, *110*, 23628.

(61) Schleyer, P. V.; Manoharan, M.; Jiao, H. J.; Stahl, F. *Org. Lett.* **2001**, *3*, 3643.

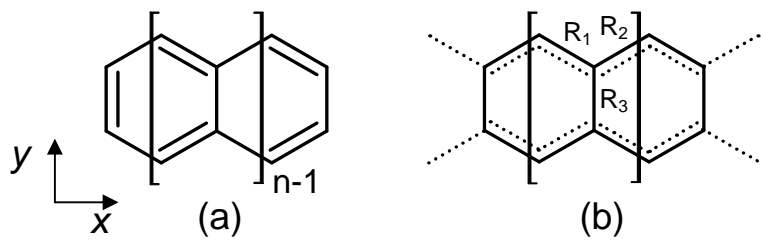

Figure 1.

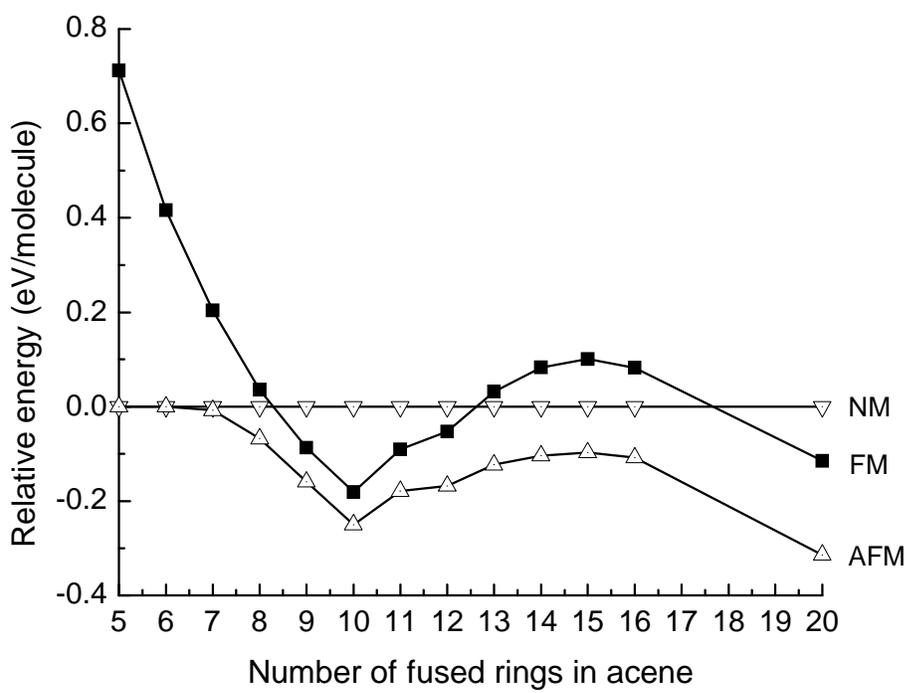

Figure 2.

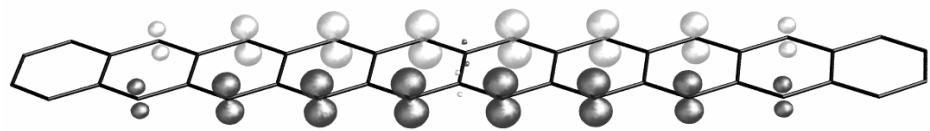

Figure 3.

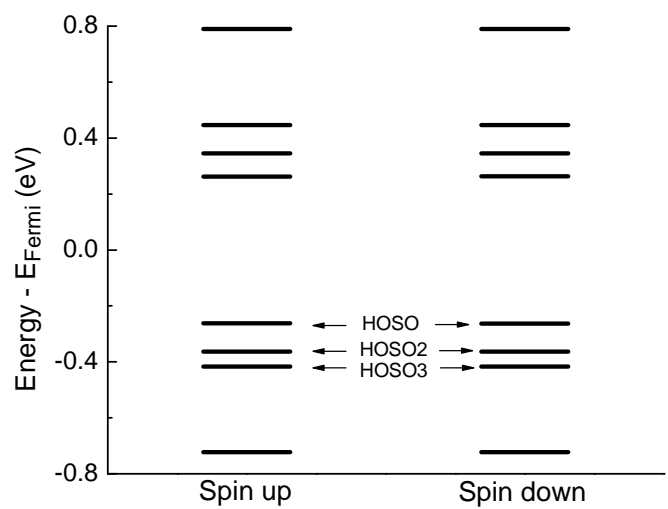

Figure 4.

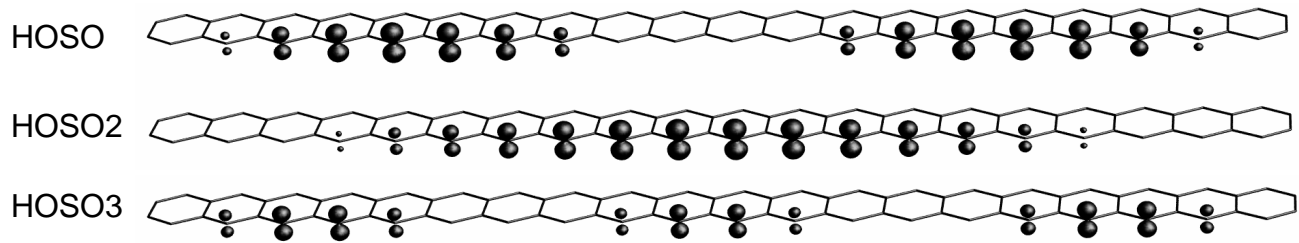

Figure 5.

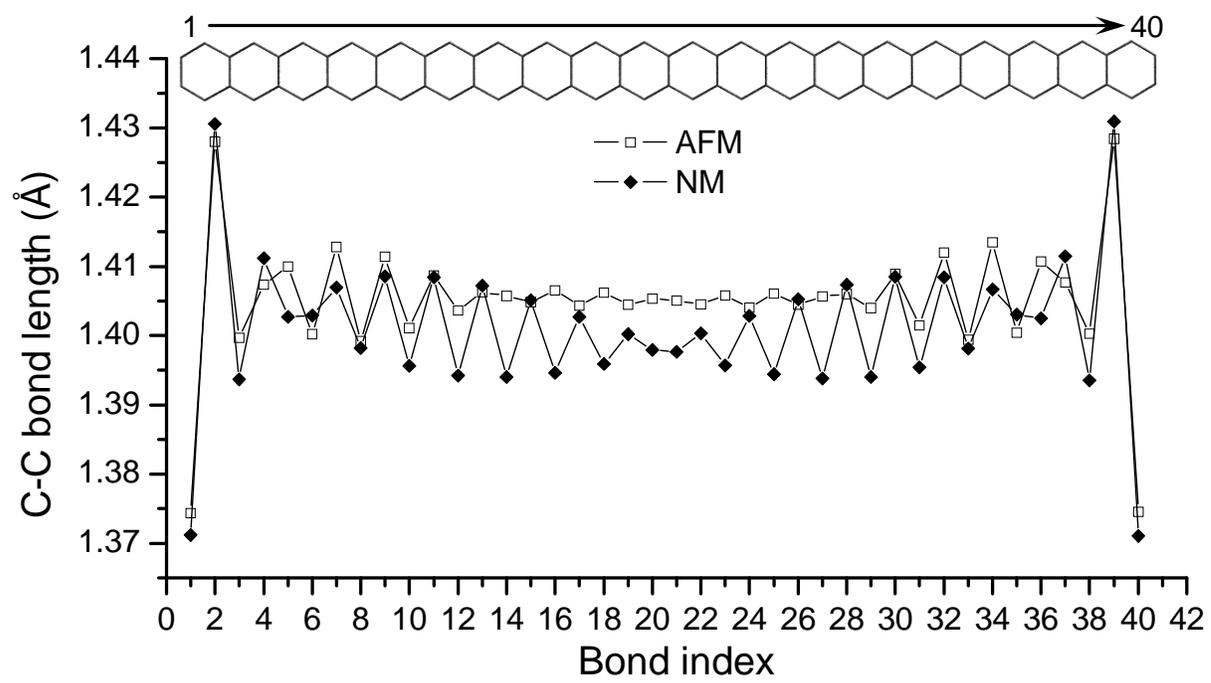

Figure 6.